\documentclass[groupedaddress,superscriptaddress,showpacs,showkeys,bibnotes,amsmath,amssymb,floatfix,reprint]{revtex4-2}
\usepackage{algorithmic}
\usepackage{graphicx}% Include figure files
\usepackage{dcolumn}% Align table columns on decimal point
\usepackage{bm}% bold math
\usepackage{hyperref}% add hypertext capabilities
\usepackage[mathlines]{lineno}% Enable numbering of text and display math
\usepackage{mathtools}
\usepackage{blkarray, bigstrut} %
\begin{document}
\preprint{TBD}

\title{Scaling Behavior of Public Procurement Activity}
\author{António Curado}
    \affiliation{NOVA Information Management School (NOVA IMS), Universidade Nova de Lisboa, Campus de Campolide, 1070-312 Lisboa, Portugal}
\author{Bruno Damásio}
    \affiliation{NOVA Information Management School (NOVA IMS), Universidade Nova de Lisboa, Campus de Campolide, 1070-312 Lisboa, Portugal}
    \affiliation{Instituto Superior de Economia e Gestão (ISEG) - Universidade de Lisboa, Lisboa, Portugal}
\author{Sara Encarnação}
    \affiliation{Interdisciplinary Centre of Social Sciences (CICS.NOVA, FCSH), Universidade Nova de Lisboa, Lisboa, Portugal}
    \affiliation{ATP-group, 2744-016 Porto Salvo, Portugal}
\author{Cristian Candia}
    \affiliation{Data Science Institute, Facultad de Ingenier\'ia, Universidad del Desarrollo, Las Condes, 7610658, Chile}
    \affiliation{Kellogg School of Management, Northwestern University, Evanston, IL 60208, USA}
    \affiliation{Northwestern Institute on Complex Systems (NICO), Northwestern University, Evanston, IL 60208, USA}
    \affiliation{Centro de Investigación en Complejidad Social (CICS), Facultad de Gobierno, Universidad del Desarrollo, Las Condes 7550000, Chile.}
\author{Flávio L. Pinheiro}
    \affiliation{NOVA Information Management School (NOVA IMS), Universidade Nova de Lisboa, Campus de Campolide, 1070-312 Lisboa, Portugal}
    \email{fpinheiro@novaims.unl.pt}

\date{\today}
\begin{abstract}
Public procurement refers to the purchase by public sector entities, such as government departments or local authorities, of Services, Goods, or Works. It  accounts for a significant share of OECD countries’ expenditures. However, while governments are expected to execute them as efficiently as possible, there is a lack of methodologies for an adequate comparison of procurement activity between institutions at different scales, which represents a challenge for policymakers and academics. Here, we propose using methods from urban scaling laws literature to study public procurement activity among $278$ Portuguese municipalities between 2011 and 2018. We find that public procurement expenditure scales sub-linearly with population size, indicating an economy of scale for public spending as cities increase their population size. Moreover, when looking at the municipal Scale-Adjusted Indicators (the deviations from the scaling laws) by contract type -- Works, Goods, and Services -- we obtain a new local characterisation of municipalities based on the similarity of procurement activity. These results make up a framework for quantitatively studying local public expenditure by enabling policymakers a more appropriate ground for comparative analysis.
\end{abstract}
\maketitle

%%%%% INTRODUCTION %%%%%
\section*{Introduction}
Public procurement contracts -- defined by the OECD as the purchase by governments and state-owned enterprises of goods and services \cite{sOECD2013} -- are an essential public sector instrument allowing policymakers to effectively push-forward inclusive socio-economic standards \cite{PublicSecProd}, promote innovation and economic growth \cite{edler2007public, uyarra2010understanding, aschhoff2009innovation} policies. Among OECD countries, public procurement weighs, on average, $29\%$ of all governmental expenditures \cite{PublicSecProd} ($14\%$ among EU countries \cite{UCPPStudy2016}) and $12\%$ of global OECD countries GDP. Moreover, given the relevance for economic activity of public procurement mechanisms, the European Commission has established a common framework for public procurement aimed at ensuring \textit{equal treatment and transparency, reduce fraud and corruption and remove legal administrative barriers to participation in cross-border tenders} \cite{PPGuide}. Not only that, but data on public procurement contracts should also constitute the basis for analytical frameworks that effectively evaluate the effectiveness and impact of public sector activities at different scales and dimensions comparatively. However, few studies have explored adequate methodologies that account for non-linearities in spending dynamics or examined the inference potential in public procurement data \cite{kristoufek2012exponential,wachs2020corruption, walker2009sustainable}.

Here, we use methods commonly used in urban scaling laws literature, rooted in statistical physics and complexity sciences, to characterise the public procurement activity of Portuguese municipalities. Urban scaling laws  \cite{bettencourt2010urban, rybski2019urban, batty2012building, gao2019computational} have been widely used across different disciplines to describe the relationship between socio-economic indicators in relation to the size of population agglomerates. The resulting literature prompted a revision of current urban planning frameworks and comparative indicators \cite{bettencourt2010unified} while leading researchers to look for universal laws in cities and urban growth. In that sense, the application of these methods to study public procurement activity can, in our view, lead to a similar revision on the current frameworks to evaluate regional public procurement policies.

In general, urban scaling laws models the relationship between an indicator, $Y$, and the size (e.g., population, area) -- $X$ -- of a set of sectional units (\textit{e.g.}, cities or urban areas) as a power-law, that is
\begin{equation}
    Y \sim \alpha X^\beta
\end{equation}
where $\beta$ is the scaling factor and $\alpha$ represents the natural baseline activity of a region \cite{molinero2019geometry,isalgue2007scaling}. In the context of urban scaling laws, several indicators -- such as, water consumption, housing, or jobs \cite{arcaute2015constructing} -- have been shown to follow a linear relationship ($\beta = 1.0$). However, the more interesting cases are those in which $Y$ exhibits a super-linear ($\beta > 1$) or sub-linear ($\beta < 1$) relationship with $X$. Such cases identify particular indicators that either scale above (super-linear) or below (sub-linear) linear growth with increasing population size. Super-linear behavior is often observed in the regional economic output \cite{lobo2013urban,schlapfer2014scaling, bettencourt2013origins, strano2016rich, van2016urban, van2019urban, balland2020complex}, energy consumption and pollution \cite{horta2010discerning, oliveira2014large,fragkias2013does}, employment \cite{bettencourt2016urban, gomez2018explaining} criminality \cite{alves2017role, gomez2012statistics, bettencourt2007growth, alves2013distance, gomez2018explaining}, number of patents \cite{bettencourt2007invention, bettencourt2007growth, bettencourt2016urban, balland2020complex}, wages \cite{bettencourt2007growth}, employment in R$\&$D \cite{bettencourt2007growth} and urbanized areas \cite{bettencourt2016urban}. Examples of Sub-linear relationships are found in the total length of road networks \cite{bettencourt2007growth, bettencourt2013origins} and power grids \cite{kuhnert2006scaling}. In other cases, like supply networks, exhibit sub or super linear behaviour depending on the industry \cite{kuhnert2006scaling, hong2020universal}, and voter turnout \cite{voterTurnoutScaling}. 

In summary, population agglomerates exhibit trade-offs between the output of human activity and the costs of maintaining human agglomerates. It is noteworthy to mention that the present analysis at the municipality unit is not directly comparable with past works in urban scaling laws literature \cite{alvioli2020administrative}, which have focused on studying cities from a functional perspective \cite{rybski2019urban,batty2008size,cottineau2017diverse,cottineau2015paradoxical,arcaute2015constructing}. Instead, and due to the local scale nature of procurement activities, we focus on the administrative boundaries of city governance in Portugal,\textit{i.e.}, namely the municipality \cite{van2020urban,altmann2020spatial}. Moreover, while past works explore how the population size of city impacts its function in different dimensions (e.g., human, cultural, and innovation outputs but also industry prevalence and infrastructure costs), our study explores an administrative dimension of cities through public expenditure.

Municipalities constitute an interesting intersection between a regional unit of governance and a population agglomerate \cite{van2020urban,altmann2020spatial}. It is also a unit at which policy related questions to procurement activity, in terms of sustainability goals and execution efficiency, are particularly relevant \cite{hyytinen2007politics,broms2017procurement,gelderman2015implementing,pavel2013auctions}. In Portugal, Municipalities are a Local Administrative Unit \cite{lauseurostat}, which form the building blocks of the NUTS (Nomenclature of Territorial Units for Statistics) European regional units, and they represent the second-largest administrative division whose governance body is elected by universal suffrage. They are also the administrative division with the most stable regional boundaries and upon which city governance responsibility falls. Thus being, a suitable candidate to study the scaling behaviour of procurement activity. 

We start with a description of the data sources and data pre-processing steps. Then, we characterised municipality procurement activity from an urban scaling law perspective, showing that procurement activity scales sub-linearly with population size. We then use Scale-Adjusted Indicators (the deviations/residuals from the scaling laws) to quantify the differences in procurement activity between different regions (groups of municipalities), but also to obtain a new regional characterization of municipality through their similarity in revealed activity. We conclude with final remarks and a discussion of future working directions. 

%%%%% DATA %%%%%
\section*{Data}
We used data on Portuguese public procurement contracts sourced from the open-access governmental portal BASE (\href{http://www.base.gov.pt}{base.gov.pt}). BASE is a public repository managed by the \textit{Instituto dos Mercados Públicos, do Imobiliário e da Construção} (IMPIC) and results from the efforts of the Portuguese government to comply with European open data policies established in 2004 \cite{EDR200418EC}. Since 2008 and by decree\cite{CPC}, public administration bodies have to publish their procurement activity online through BASE \cite{costa2013evidence}.

\begin{figure*}[!th]
	\centering
	\includegraphics[width=\textwidth]{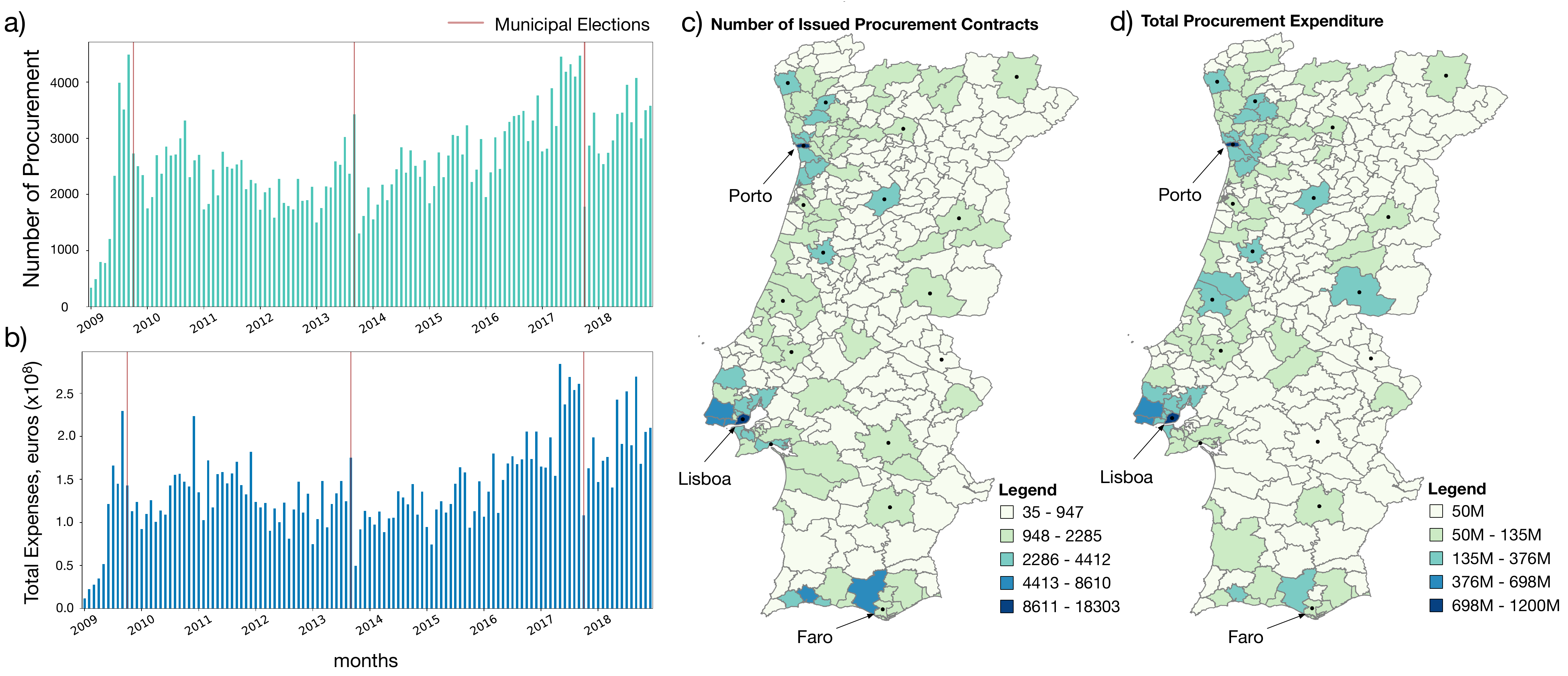}
	\caption{
	    a) Number of monthly procurement tenders issued between 2009 and 2019 by Portuguese municipalities. b) Total value in euros derived from procurement, monthly issued by Portuguese municipalities between 2009 and 2019. In both a) and b), each bar corresponds to a month/year, and vertical red lines indicate Municipal elections held nationwide. c) Spatial distribution of the total number of issued procurement tenders by a municipality. d) Spatial distribution of the total value spent in procurement contracts per municipality, between 2009 and 2019.
	\label{Figure1}}
\end{figure*}

The working data set comprises $930,513$ contracts issued between January 2009 and December 2018. Each contract relates an issuer that acquires services/goods/works from a supplier and contains information on the issue date; the value of the contract (in euros); type of contract; and Fiscal Numbers of both the issuer and supplier. We analysed contracts issued by the $278$ Municipalities that constitute mainland Portugal\footnote{We did not consider municipalities in the Azores and Madeira archipelagos as they represent autonomous administrative regions.}. Since municipalities can also constitute municipal firms (i.e., a municipality can be the single shareholder of another firm), we aggregated all municipalities and respective child firms into a single entity. The aggregation was manually-curated with support from the Annual Financial Booklet of Portuguese Municipalities  \cite{carvalho2008anuario, carvalho2013anuario, carvalho2014anuario}. The pre-processing steps include:
\begin{enumerate}
    \item Removing observations with contract values equal or smaller than one;
    \item Identifying the Fiscal Number of each Municipality to use them as a primary key;
    \item Aggregating municipal firms to the parent Municipality;
    \item Discarding all non-Municipality related procurement contracts;
    \item The value of contracts that involved more than one municipality were split equally among all participating municipalities (the issuers).
\end{enumerate}
The final dataset comprises $310,819$ contracts totalling a value of $16.9$ Billion Euros. Panel a) of Figure~\ref{Figure1} shows the monthly number of contracts issued, while panel b) shows the total value of those contracts. Through visual inspection, it is possible to identify a tendency for municipalities to increase the number of procurement contracts issued in the months leading to elections (red vertical lines in Figure~\ref{Figure1}a). However, the same does not necessarily translate into an increase in expenditure. Figure~\ref{Figure1}c) and d) show the spatial distribution of the total number of contracts (c) and the total value per municipality (d). 

Each procurement is associated with the type of contract it represents, which follows the standard classification from the European Commission \cite{PPGuide}: Work contracts designate contracts whose execution and/or design include civil engineering works such as roads or sewage plants; Goods contracts identify contracts that are associated with the purchase, lease, or rental of products such as vehicles or computers; Service contracts involve all contracts that have as an object the provision of services such as consultancy, training, or cleaning services. Hence, we computed the total expenditure in procurement contracts per year for each municipality, as well as the total expenditure by procurement contract type. Moreover, since the annual expenditure was rather noisy, we applied a sliding window technique (moving average) of three years. In that sense, the procurement values at year $t$ correspond to an average of the values from years $t-2$, $t-1$, and $t$. The reported noise can have multiple sources. For instance, a municipality might have issued a procurement for the execution of construction in one year that was reflected in the forthcoming annual budgets and thus decreased its construction activity in the following years.  

Finally, we enriched the data set with additional indicators by municipality and year. From Pordata \cite{pordata} we sourced data on Social Integration Income; House Prices; Number of Public Workers; Total Births; Number of Large Corporations; Number of Divorces; Amount of Credit; Number of Medical Doctors; Number of Culture Attendees; Imports and Exports Volumes; and Environment Expenses. While from INE \cite{ine19} we sourced ATM Withdrawals, Municipal Property Tax, Volume of Business in Accommodation, Catering, and Retail; Individual Gross Income; Average Salary of Full-Time Workers. We use these indicators to provide a point of reference to the analysis of public procurement activity, while allowing to compare with previous urban scaling laws literature.

%%%%%%%%%%%%%%%%%%%%%%%%%%%%%%%%%%%%%%%%%%%%%%%%%%%%%%%%%%%%%%%%%%%%%%%%%%%%%%%%%%%%%
%%%%%%%%%%%%%%%%%%%%%%%%%%%%%% Results and Discussion %%%%%%%%%%%%%%%%%%%%%%%%%%%%%%%
%%%%%%%%%%%%%%%%%%%%%%%%%%%%%%%%%%%%%%%%%%%%%%%%%%%%%%%%%%%%%%%%%%%%%%%%%%%%%%%%%%%%%

%%%%% Scaling Laws %%%%%
\section*{Results and Discussion}
\subsection*{Scaling Laws of Municipal Procurement Expenditure}
\begin{figure*}[!th]
	\centering
	\includegraphics[width=\textwidth]{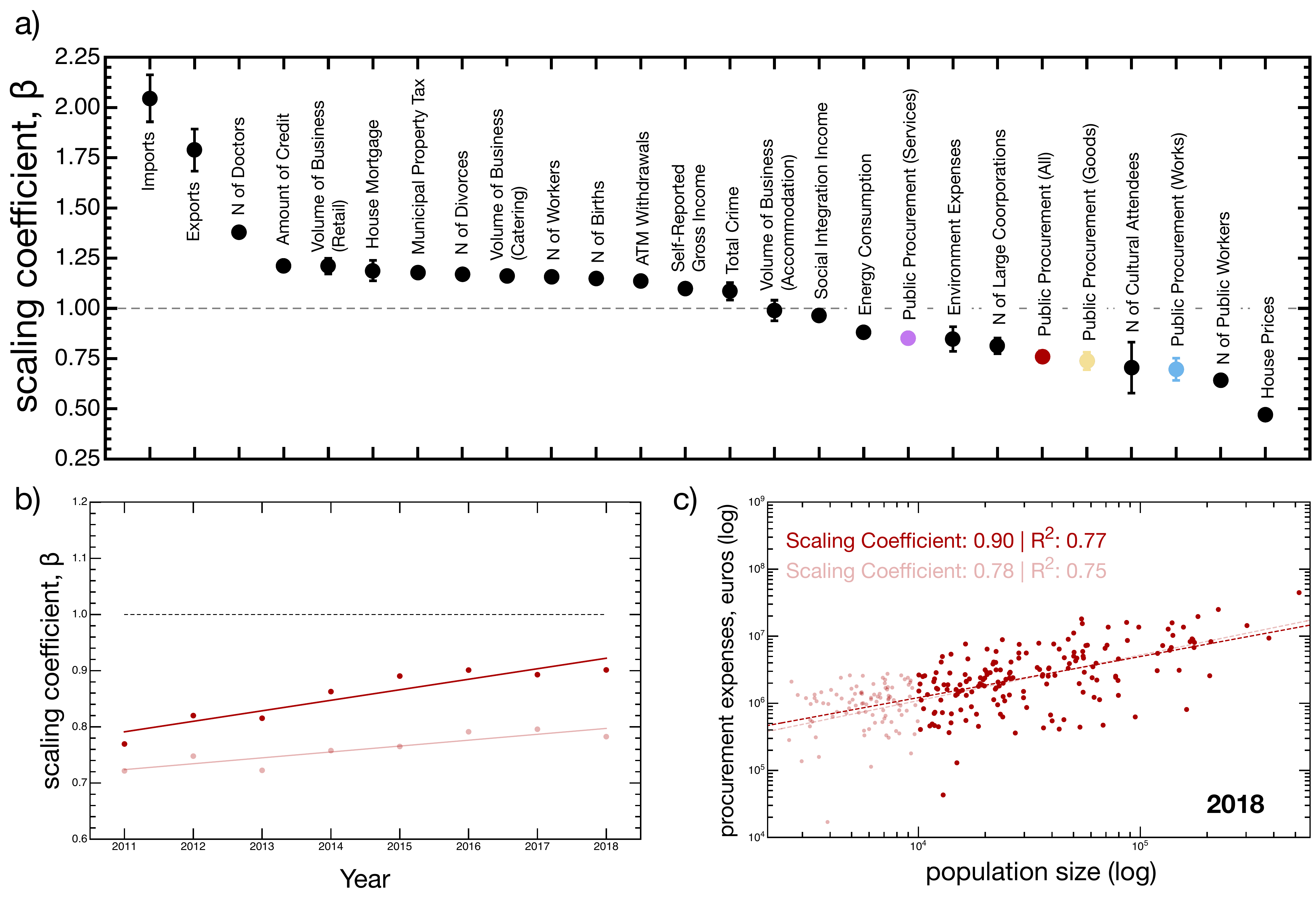}
	\caption{
        a) Average scaling coefficients for multiple socio-economic metrics (dark) and all procurement expenses (red). Moreover, we also show the coefficients obtained for the different types of procurement contracts: works (blue), goods (yellow), and services (purple). Error bars indicate the standard deviations of estimated coefficients for the different years. Panel b) shows the annual changes in the scaling coefficients for procurement activity. Panel c) exemplifies the identified relationships between the procurement expenditure (euros) and population size for the year of 2018. In panels b) and c) we highlight the results for municipalities with more than $10^4$ residents in dark red, light red indicates the results obtained when considering all municipalities.  
	\label{Figure2}}
\end{figure*}

We start by comparing the estimated scaling coefficients from municipal procurement activities with those estimated from an extensive set of socio-economic indicators ( Figure~\ref{Figure2}a). The coefficients were estimated independently for each year between 2011 and 2018. Figure~\ref{Figure2}a shows the average coefficient (Y-Axis) per indicator (X-Axis) with error bars representing the standard deviation. 

In general, the obtained scaling coefficients are inline with previous findings in the urban scaling laws literature, thus supporting the choice of analysis at the municipality level. Namely, a super-linear behaviour was observed for the volume of imports ($\beta = 2.05 \pm 0.12$) and exports ($\beta = 1.79 \pm 0.10$), number of medical doctors ($\beta = 1.38 \pm 0.01$), volume of business from retail except car sales ($\beta = 1.21 \pm 0.01$), amount of credit ($\beta = 1.21 \pm 0.04$), municipal property tax collected  ($\beta = 1.19 \pm 0.05$), number of divorces ($\beta = 1.18 \pm 0.02$), total volume of house mortgages ($\beta = 1.17 \pm 0.02$), volume of business from catering ($\beta = 1.16 \pm 0.03$), number of workers ($\beta = 1.16 \pm 0.01$), number of births ($\beta = 1.15 \pm 0.02$), ATM withdrawals ($\beta = 1.14 \pm 0.01$), self-reported gross income ($\beta = 1.10 \pm 0.01$), and reported crime ($\beta = 1.08 \pm 0.04 $). Linear scaling was observed for total volume of business from accommodation  ($\beta = 0.99 \pm 0.05$). Sub-linear scaling was observed for energy consumption ($\beta = 0.88 \pm 0.006$), social integration income ($\beta = 0.97 \pm 0.03$), environment expenses ($\beta = 0.84 \pm 0.06$), number of large corporations ($\beta = 0.81 \pm 0.04$), number of culture attendees ($\beta = 0.70 \pm 0.13$), number of public workers ($\beta = 0.70 \pm 0.06$), and house prices ($\beta = 0.47 \pm 0.02$).

Figure~\ref{Figure2}b--c explores the results obtained from the total procurement expenses per municipality in more detail. Figure~\ref{Figure2}b shows the yearly change in the scaling coefficient, which exhibits an upward temporal trend. Light-coloured points indicate scaling coefficients estimated when considering all municipalities, while dark-coloured ones only consider municipalities with a population size larger than  $10^4$. Figure~\ref{Figure2}c shows a representative example of the scaling behavior from the year 2018. The threshold was set to filter out low populated municipalities with small procurement activity, and does not reflect any underlying administrative differences between municipalities (we discuss more in detail the sensibility of our results to the choice of threshold below). In that sense, Figure~\ref{Figure2}b and \ref{Figure2}c show the impact of including (lighter color) or not (darker color) municipalities with a population lower than $10^4$. In all cases, the coefficient shows a sub-linear relationship between the total public procurement expenditure and population size.

\subsection*{Scaling of Procurement Activity by Contract Type}
\begin{figure*}[!t]
	\centering
	\includegraphics[width=\textwidth]{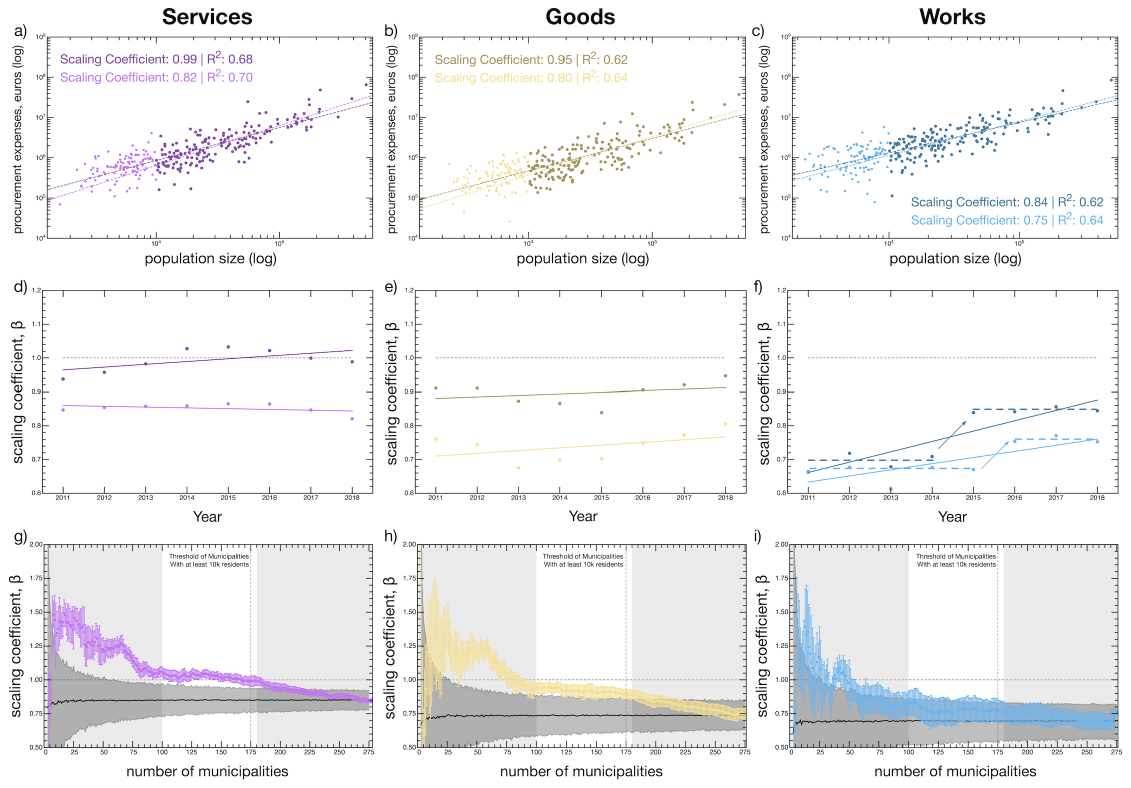}
	\caption{
	    Decomposition of the scaling factors evolution per procurement contract type.
	    Panels a–c) Relationship between total procurement expenses, by type, in 2018. Panels d–e) Scaling coefficient per year for each contract type. 
	    In panels a–f) lighter colours indicate the analysis conducted on all Municipalities, darker colours on the subset of municipalities with a population greater or equal to $10^4$. In Panels d–f) lines indicate the best linear model but should only serve as a guideline as they are not statistically significant at p-value threshold of $0.1$.
	    Panels g–i) Robustness checks on the estimated scaling coefficients. Colored points show the estimated coefficient when different population thresholds are applied (we do so by adding the least populated municipalities, from the left to the right side, to the sample). Error bars indicate the standard deviation of the estimated scaling coefficient. Black line indicate the average coefficient from bootstrapping the estimation from X municipalities selected at random with replacement (dark shaded area show the $95\%$ confidence interval). Vertical dashed line indicates the point at which added municipalities have a population lower than $10^4$.
	\label{Figure3}}
\end{figure*}

Figure~\ref{Figure3} extends the analysis done in Figure~\ref{Figure2}b and Figure~\ref{Figure2}c to different procurement contract types: Services, Goods, and Works. Like in Figure~\ref{Figure2}, Light colours refer to the entire set of municipalities, while darker colours represent the sample of municipalities with a population size larger than $10^4$. Figure~\ref{Figure3}a--c shows the scaling relationships in the year of 2018 for all three types of procurement contracts. As with the results in Figure~\ref{Figure2}c, Goods and Works contracts show a clear sub-linear scaling. In contrast, Services show an almost linear relationship if only the most populated municipalities are considered, but a sub-linear relationship is recovered when the entire set is under consideration. 

Focusing in the coefficients obtained for the entire set of municipalities. Public procurement contracts are associated with the costs of a city governance: maintenance, expansion, and functioning of public infrastructures. In other words, it corresponds to an infrastructure cost. In that sense, the sub-linear behavior is, thus, inline with past results from urban scaling laws literature. Moreover, the sub-linear behavior suggests that we are in presence of an economy of scale. That is, larger municipalities need to spend less \textit{per capita} through public contracts to perform their activities than smaller municipalities. Another way of thinking about the sub-linearity is that, in a \textit{per capita} basis, for each euro spent by one municipally another that is x times larger would spends only a fraction that value.

Moreover, in Figure~\ref{Figure3}d--f, we show that, unlike the results in Figure~\ref{Figure2}b, the annual upward trend of the scaling coefficient is absent in Goods and Works contracts (note that the slope coefficient in all trend lines are non-significant at p-value threshold of 0.1). However, in the case of Works contracts, these exhibit a transition, around 2014/16, between two seemingly stable regimes, a behaviour that contributes to understand the upward trend identified in Fig.~\ref{Figure2}b. The transition observed in the Works contracts' scaling coefficient implies that, after 2014/15, the most populated municipalities started to spend proportionally more than less populated ones, albeit the persistence of a sub-linearity character still means that less populated municipalities spend more in a \textit{per capita} basis. But what can explain such abrupt change in the coefficients of the Works contracts?

Between 2013/15 several events took place that can help us better understand the context in which the above-mentioned transition took place. Portugal left the Troika bailout program in 2014 \cite{pereira2015portugal}. The program is well known for having introduced unpopular policies to control public sector finances. Moreover, the period is marked by two nationwide elections -- municipal elections of 2013 and the parliamentary elections on 2014 -- that lead a shift in the political landscape from center-right to center-left \cite{hardiman2019tangling,fernandes2017late,costa2020does}. Arguably, such change could have contributed to a shift in the philosophy of public investments. Economically, 2014 is marked by an increase in private sector activity accompanied by a steadily raise in tourism flows to Portugal\cite{santos2020tourism,fuinhas2019economic}. Although contextually important, it is difficult to trace route these events to the underlying cause of the transition in the scaling coefficients of Works contracts. Instead, they help to explain the steady decline in procurement activity during the bailout program, and the observed rise afterwards (see Figure~\ref{Figure1}a and \ref{Figure1}b), as it become easier for public administration to obtain funding.

However, a more significant event took place in 2014. A new municipal finances law, which entered into practice in January 1st 2014, tightened the ability of municipalities to contract debt. Interestingly, at the time, it was widely speculated that the new law would put a particular pressure on the small municipalities ability to finance their investments \cite{tonelottopublic, pinto2015implicaccoes, joao2014estudo, dos2020sustainability, piresdeterminants}. Naturally, such uneven pressure could have lead to the behaviour observed in Figure~\ref{Figure3}f: a decrease in the procurement activity inversely proportional to the population size increases the slope of the relationship. It is also reasonable to understand that such law would manifest particularly in some contract types rather than others: Works contracts are linked with construction projects and public investment in infrastructure; while Goods and Services contracts that are associated with the normal functioning expenses of the municipality (e.g., engineering services, finance and accounting services, training and development, furniture, IT equipment, books, vehicles, medical supplies and other commodities) where cuts are less likely to occur.

We conclude this section by testing the robustness of the scaling coefficients by setting different population thresholds and by bootstrapping the coefficients' estimation. 

Figure~\ref{Figure3}g--i explores the impact in the estimated scaling coefficient (Y-axis) by considering only the $n^{\textit{th}}$ (X-axis) most populated municipalities (colored points with error bars) or by considering a random sample of $n$ (X-axis) municipalities (black curve)\footnote{Results correspond to the average from $500$ independent samples. Each sample is done by first selecting a year at random, and then $n$ municipalities also at random.}. When performing a threshold by population size, we observe three distinct regimes: First, when only the most populated municipalities are considered (left-hand shaded area Figure \ref{Figure3}g--i), there is a high variance in the estimated scaling coefficients; Second, an intermediate regime in which the coefficient remains stable to variations in the number of municipalities (center white area Figure \ref{Figure3}g--i); Third, a regime of linear convergence of the coefficient (right-hand shaded area Figure \ref{Figure3}g--i). In the second, and stable, regime we qualitatively observe the same relationship between the estimated coefficients of the three types of contracts: Services with higher coefficient close to one while Works has the lowest coefficient around $0.75$. Also important to note, the threshold of $10^4$ inhabitants is located in the right-most boundary of stable regime ($x = 175$, vertical dashed line). Moreover, these regimes that appear when filtering the municipalities by population size, are absent when the coefficients are estimated by performing a random samples of similar size, where we recover the coefficient estimated from the entire set of municipalities (black curve in Figure~\ref{Figure3}g--i).

Next, we random sample $n$ (with $n$ from $3$ to $275$, Figure~\ref{Figure3}g--i) municipalities with replacement, so municipalities can be considered more than once in each random sample, and estimate the scaling coefficient of the sample. We repeat this procedure $1,000$ times for each value of $n$. Results are shown as black curve in Figure~\ref{Figure3}g--i, with confidence intervals in shade.

In sum, different population size thresholds can lead to differences in the estimated coefficients of Services, Goods, and Works (see Figure~\ref{Figure3}d--f and \ref{Figure3}g--i), yet, our results are robust and unbiased to random samples of different sizes, meaning that the data points are equally distributed in the entire domain of analysis. Moreover, since our goal is to obtain a comprehensive picture of the regional dynamics of Portuguese municipalities, one that is relevant to assess policy implications of procurement activity, we focus the analysis on the entire set of municipalities. 

%%%%% SAI %%%%%
\subsection*{Municipal Procurement Scale-Adjusted Indicators}
One major challenge when developing a comparative analysis of regional data relates to how an indicator scales along the dimension of analysis (e.g., region area, population size, etc.). For instance, it is common to compare regions on a \textit{per capita} basis. However, such comparison relies on the implicit assumption that indicators scale linear with population size, therefore the estimators will be inconsistent under sub or super-linearity regimes, in which case outcomes can suffer from increasing/decreasing returns with the population size. Since that is not the case for most indicators, see Figure~\ref{Figure2}, we can end up with erroneous conclusions. In that sense, urban scaling laws literature proposes using the residuals of each region from the specific scaling law as a reference model.

We follow by estimating the so-called \textit{Scale-Adjusted Indicators} (SAI) \cite{bettencourt2010urban,alves2013distance,alves2015scale} to quantify deviations of each municipality procurement activity from the scaling reference model. The SAI correspond to the residuals, which are computed as
\begin{equation}
    \textit{SAI}_{i,t} = log_{10}\frac{Y_{i,t}}{Y(N_{i,t})}
\end{equation}
where $Y_{i,t}$ is the observed expenditure of municipality $i$ on year $t$, and $Y(N_{i,t})$ is the predicted value given the population size of such municipality. Unlike \textit{per capita} indicators, the SAI are dimensionless and independent of population size \cite{bettencourt2010urban,alves2013distance,alves2015scale}. The SAI capture human and social dynamics specific to a given place and time, allowing for a population-unbiased comparison between regional administrative bodies. For instance, SAI have been used to identify clusters of regions with similar activity patterns \cite{bettencourt2010urban}.

\begin{figure*}[!th]
	\centering
	\includegraphics[width=\textwidth]{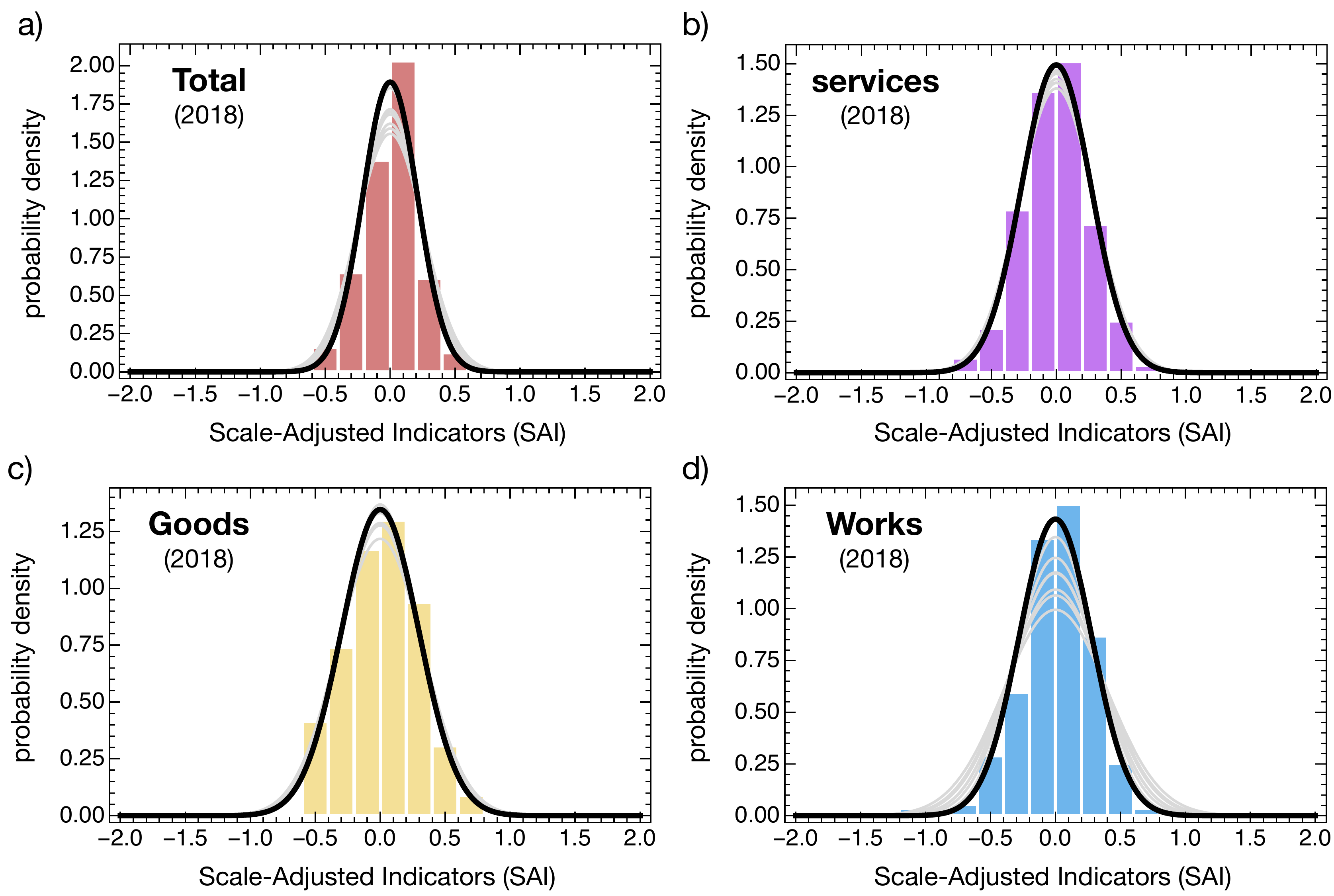}
	\caption{
	    Distributions of Scale-Adjusted Indicators (SAI). Bars show the distribution of SAIs for 2018; curves show the best fitted Normal Distribution to the SAI data for each year. Except for the observations for one year and one type of contract (Goods in 2014) the hypothesis that SAI follow a normal distribution cannot be disproved using Cramér–von Mises criterion at the p-value = 0.05 threshold. .
	\label{Figure4}}
\end{figure*}

Figure~\ref{Figure4} shows the distribution of SAI obtained for different types of contracts for 2018 and the best fit Normal Distribution for the SAI estimated per year. In all but one case, SAI are Normally Distributed. Moreover, the SAI are uncorrelated with population size and show no heteroscedasticity.

\subsection*{Regional Divide}
One application of Scale-Adjusted Indicators is to identify and compare patterns between groups of regions \cite{bettencourt2010urban}. By helping to identify commonalities but also exceptions, they can aid policymakers in pushing adequate policies for regional development \cite{pumain2019two}. In that sense, Portugal is marked by different regional development profiles related to the natural, social, and economic diversity that loosely runs from North to South and its urban system anchored on two main metropolitan areas (Lisbon and Porto) \cite{onnerfors2019eurostat}. On the other hand, the country deals with substantial challenges stemming from significant migration movements towards coastal regions (and the ongoing decline in population growth) that, in time, amplified territorial disparities mainly marked by the dichotomy of Interior/Coastal regions \cite{alegria1990norte,rees1998internal,ferrao2002portugal,PNPOT2018,PNPOT2019,nordhaus2011geographically,nordhaus2011geographically}.

\begin{figure*}[!th]
	\centering
	\includegraphics[width=\textwidth]{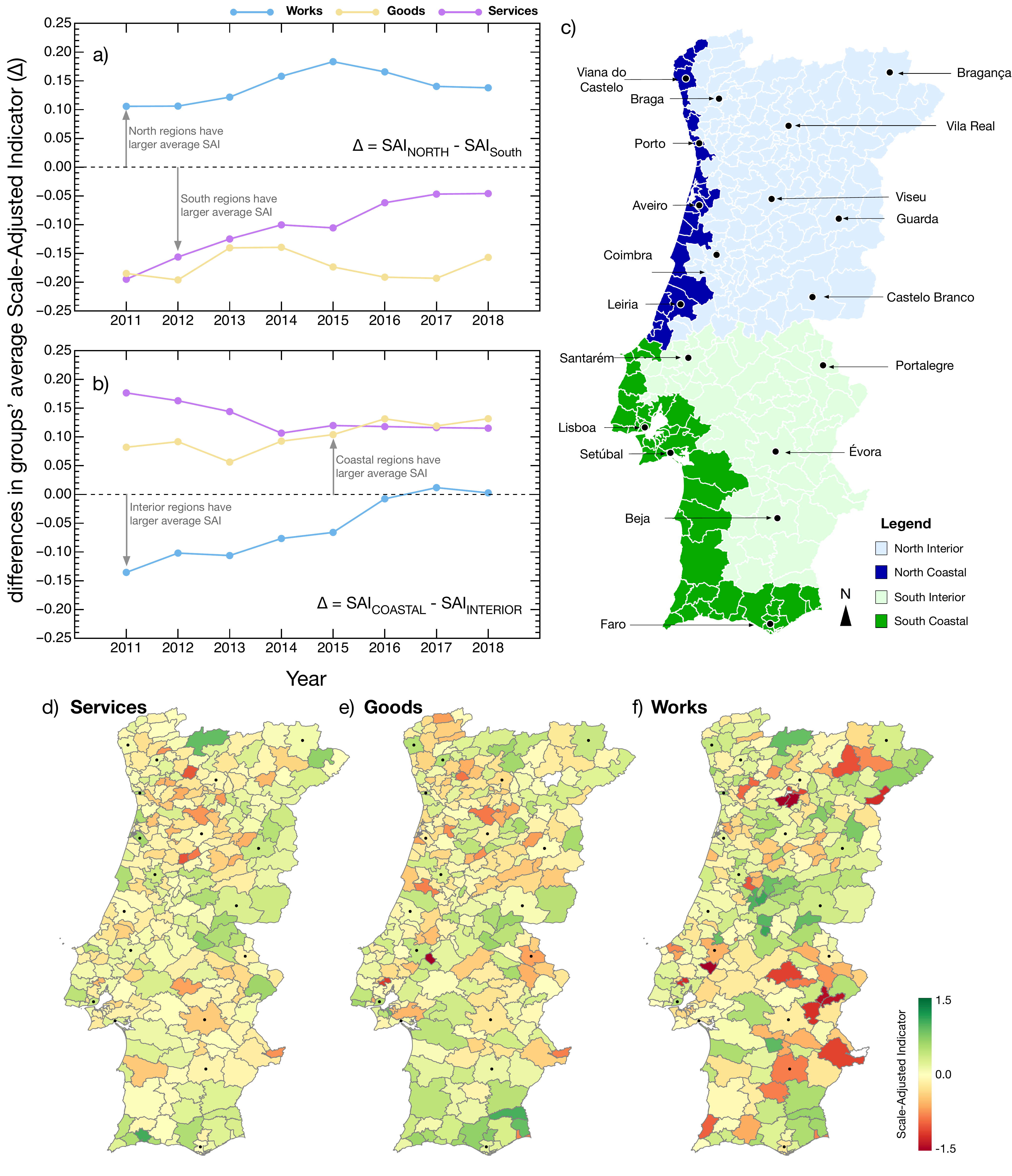}
	\caption{
	    Annual Differences in the Scaling-Adjusted Indicators between different groups of Portuguese Municipalities. Panel a) shows the differences between North and South, and Panel b) shows the differences between Coastal and Interior. Panel c) shows the Portuguese municipalities coloured according to the groups to which they were assigned. District capitals are indicated. Panels d–f show the Scale-Adjusted Indicators of each municipality for each procurement contract type in the year of 2018. Minimum and maximum of the colour range are set to the maximum absolute Scale-Adjusted Indicator observed. 
	\label{Figure5}}
\end{figure*}

In light of such regional realities in Portugal, we investigate whether public procurement activities  exhibit different and distinguishable patterns. We grouped municipalities according to whether they are located on the Coast/Interior or in the North/South\footnote{It is noteworthy to mention that there is no generally agreed upon definition of these regional groups.}. We defined as Coastal municipalities all those that have a coastline or that are enclaves of municipalities with a coastline, else they were categorised as Interior. Moreover, we used the coordinates of each municipality city center as a point of reference to classify them as being in the North or South. In particular, we classified as North the 140 municipalities whose city coordinates are the northernmost; the remaining 138 were classified as being in the South. Figure~\ref{Figure5}c shows the classification of each municipality.

Figure~\ref{Figure5}a compares the average SAI between North and South municipalities ($\Delta = \text{SAI}_{\text{NORTH}}-\text{SAI}_{\text{SOUTH}}$) and Figure~\ref{Figure5}a repeats the analysis between Coastal and Interior municipalities ($\Delta = \text{SAI}_{\text{COASTAL}}-\text{SAI}_{\text{INTERIOR}}$). The differences in SAI between the groups -- Northern/Southern \cite{onnerfors2019eurostat} or Coastal/Interior \cite{nordhaus2011geographically} -- reveal, and help, to quantify the existence of clear dichotomies that characterize the common perceptions and differences between regional groups and their governance (see Figure~\ref{Figure5}a and \ref{Figure5}b). For instance, Northern municipalities tend to exhibit, in average, larger SAI in Works, while Southern Municipalities are characterised by larger SAI in Services and Goods. These patterns remain qualitatively the same over the years. Moreover, Coastal municipalities tend to have higher SAI in Goods and Services procurement contracts, while Works contracts have evolved from being larger in the Interior to reach parity since 2016. Figure~\ref{Figure5}d--f show the spatial distribution of the 2018 SAI across the Portuguese municipalities for each type of procurement contracts. Interestingly, the distribution of SAI does not directly maps into the regional groups (North/South and Coastal/Interior), instead, they show a richer spatial distribution of procurement patterns.

Here, we propose to use the SAI to identify clusters of municipalities with similar procurement activity patterns. Municipalities were clustered using the K-Means algorithm, identifying four clusters with similar activity patterns. We performed the clustering by considering that each observation consists of the SAI for each type of procurement per municipality per year. See the Supplementary Information for more information about the clustering procedure.

\begin{figure*}[!th]
	\centering
	\includegraphics[width=\textwidth]{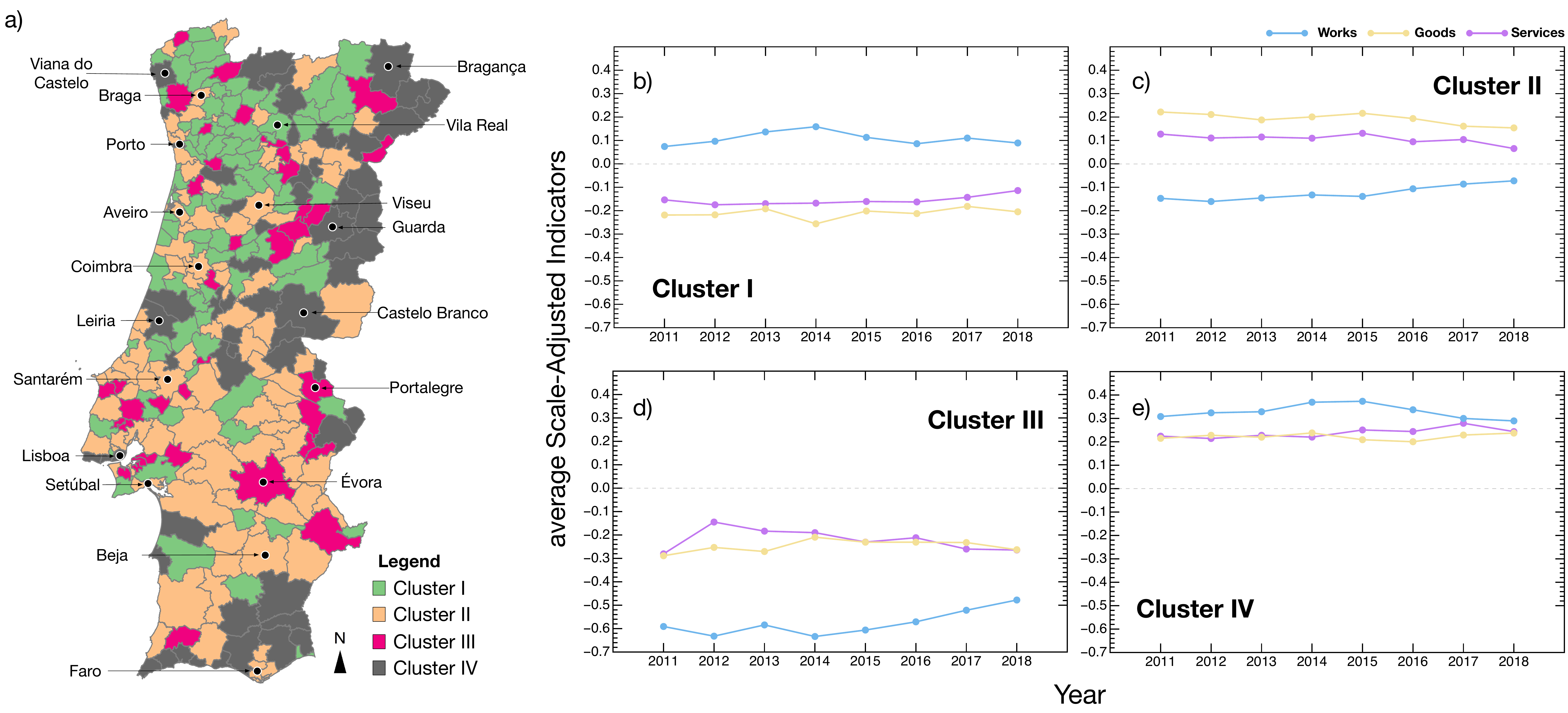}
	\caption{Clustering of Municipalities according to the Scale-Adjusted Indicators patterns.
	Panel a) associates municipalities to a cluster. Panels b--e) show the average SAI in the different procurement contract types -- Works (blue), Services (purple), and Goods (yellow) -- allowing to characterize the procurement activity patterns of each cluster.
	\label{Figure6}}
\end{figure*}

We refer to the identified clusters as I, II, III, and IV. Figure~\ref{Figure6}a shows the spatial distribution of clusters estimated from procurement activity for 2018. It is important to note that this approach has the advantage of not assuming pre-conceived perceptions of the historical, geographical, or demographic factors and translates the information revealed by (i.e., embedded in) the municipalities’ procurement activity. 

The identified clusters break the spatial homogeneity of the regional divisions studied above (North/South and Coastal/Interior). Profiling the clusters enables to reveal common patterns and similarities in governance within groups but also the differences between groups. We note that clusters I and II are characterized by geographical, political, and industrial differences. For instance, cluster I is mainly concentrated in the North (dominated by an electoral preference for center-right parties) and Cluster II in the south (dominated by left and center-left political parties) \cite{costa2020does}. These regions are also known to have different industry structures \cite{onnerfors2019eurostat}: The North being more manufacturing intensive, while the south is more Agriculture intensive. These two clusters from the archetype basis of procurement activity we would expect to observe in Portugal given its regional dichotomies.

The two remaining clusters unveil differences stemming from policy and financially related constrains. Cluster III is constituted by municipalities with a higher debt \textit{per capita} than the average (see SI), which helps to explain their consistently lower public procurement activity. In contrast, cluster IV comprises a set of municipalities where independent parties are more likely to prevail (political electoral preference is not so clear) and that have higher funds from European Union in a \textit{per capita} basis (see SI). Hence, it is not surprising their ability execute more procurements than expected in all types of contracts. These clusters also span across the entire country.

These findings may signal the presence of self-reinforcing mechanisms (Cluster I and II), that is, by means of rooted cultural and context-specific patterns of public procurement activity. However, a more robust analysis of the factors underlying such differences is needed to adequately profile each cluster and identify the socio-cultural, and political factors driving the different procurement activity profiles. Nevertheless, it is clear that SAI configures a momentous tool in what concerns the identifications of economic, political, and regional (dis)similarities and imbalances, offering a framework to adequately compare different regions.

\section*{Conclusions}
Having the correct methodology to perform comparative regional analysis of procurement activity is critical to evaluate the effectiveness of public policies and their socio-economic consequences. However, it is a challenge to develop accurate measures when indicators do not scale linearly with, for example, population size. Here, we proposed using methods from urban scaling laws to analyse procurement activities among 277 Portuguese municipalities and by contract type.

We characterised the scaling coefficient of procurement activity and put it at a glance with an array of other indicators (Figure \ref{Figure2}). Municipal procurement activity tends to scale sub-linearly with population size, meaning that increasing the population size lowers the value spent \textit{per capita} in public contracts. Such behaviour is true for both the total value spent in public procurement and among the different types of contracts (Figure \ref{Figure2} and \ref{Figure3}). 

We observe an upward trend in the annual variation of the scaling coefficients. However, we argue that such trend is being modulated by the observed transition in the Works procurement contracts, that jumped from  $\approx 0.7$ to $\approx 0.85$ between 2014/16. We link such transition to several socio-political events that took place in Portugal but, more importantly, by the new law of municipal finances that entered in practice in January 1st of 2014.

Finally, looking at the deviations from the scaling laws (the SAI), we compare differences in procurement activity between regional groups of interest, North/South and Coastal/Interior. But also we propose using the SAI to identify groups of municipalities with similar procurement activity. We showed that the resulting groups/clusters of municipalities present different procurement activity, which are rooted in their electorate political preferences and industrial structures (Cluster I and II), but also defined by financial constrains (Cluster III are reveal indebted municipalities) or policy driven (Cluster IV are revealed to received higher EU funds). These results may signal self-reinforcing mechanisms, but also potential handicaps for regional convergence and development. We believe that these findings are worth exploring in further research (\textit{e.g.}, looking for additional factors to allow a better understand the cluster stemming from public procurement activity) and in better understand and compare the national dynamics with that of other countries.

Although previous works have analysed procurement data  \cite{kristoufek2012exponential}, they have done so using relatively smaller samples. Ongoing work aims at developing a more robust model to understand the link between public procurement activity at the regional level and economic development. A challenge that requires identifying the appropriated indicators and the adequate model specification \cite{keuschnigg2019urban}. Moreover, future research may also extend this analysis to other countries and regions, particularly European public procurement repositories. The latter would allow us to validate the identified similarities and differences in behaviours across countries with different processes and cultures. 

\section*{Acknowledgments}
The authors are thankful to Dominik Hartman, Fernando Bação, and Lia Quadros Flores for the useful discussions, and to Filipe Freire for helping in the initial data curation. B.D. acknowledges the research protocol established with \textit{Instituto Nacional de Estatística} (INE) that provided access to regional data. F.L.P and B.D. acknowledge the financial support provided by FCT Portugal under the project UIDB/04152/2020 -- Centro de Investigação em Gestão de Informação (MagIC). S.E. acknowledges the financial support of FCT and ESF, for project grant SFRH/BPD/1169337/2016 and FCT -- Foundation for Science and Technology, I.P., within the scope of the project UIDB/04647/2020 of CICS.NOVA -- Centro Interdisciplinar de Ciências Sociais da Universidade Nova de Lisboa.

\bibliography{References}
\end{document}